# Preserving Data Privacy for ML-driven Applications in Open Radio Access Networks

Pranshav Gajjar
*NextG Wireless Lab*
George Mason University, USA
Email: pgajjar@gmu.edu

Azuka Chiejina
*NextG Wireless Lab*
George Mason University, USA
Email: achiejin@gmu.edu

Vijay K. Shah
*NextG Wireless Lab*
George Mason University, USA
Email: vshah22@gmu.edu

***Abstract*—Deep learning offers a promising solution to improve spectrum access techniques by utilizing data-driven approaches to manage and share limited spectrum resources for emerging applications. For several of these applications, the sensitive wireless data (such as spectrograms) are stored in a shared database or multistakeholder cloud environment and are therefore prone to privacy leaks. This paper aims to address such privacy concerns by examining the representative case study of shared database scenarios in 5G Open Radio Access Network (O-RAN) networks where we have a shared database within the near-real-time (near-RT) RAN intelligent controller. We focus on securing the data that can be used by machine learning (ML) models for spectrum sharing and interference mitigation applications without compromising the model and network performances. The underlying idea is to leverage a (i) *Shuffling-based learnable encryption technique* to encrypt the data, following which, (ii) employ a custom *Vision transformer (ViT)* as the trained ML model that is capable of performing accurate inferences on such encrypted data. The paper offers a thorough analysis and comparisons with analogous convolutional neural networks (CNN) as well as deeper architectures (such as ResNet-50) as baselines. Our experiments showcase that the proposed approach significantly outperforms the baseline CNN with an improvement of 24.5% and 23.9% for the percent accuracy and F1-Score respectively when operated on encrypted data. Though deeper ResNet-50 architecture is obtained as a slightly more accurate model, with an increase of 4.4%, the proposed approach boasts a reduction of parameters by 99.32%, and thus, offers a much-improved prediction time by nearly 60%.**

## I. Introduction

Wireless communication networks have experienced rapid growth with the integration of machine learning (ML) that aims to apply data-driven approaches to optimize the performance of these networks [1]. The utilization of ML techniques, particularly Deep Learning (DL) algorithms extends across various layers of wireless networks encompassing the physical layer up to the application layer, covering spectrum sharing, resource management, networking, mobility management, and localization [1]. These ML and DL applications within the wireless domain often rely on shared data environments such as in the case of spectrum sharing in Citizen Broadband Radio Service (CBRS) band [2], [3]. The existence of multiple stakeholders with unrestricted access to data in such systems poses a potential risk of privacy or security breaches.

A prominent case study for a shared data environment in cellular networks can be the paradigm of Open Radio Access Network (O-RAN), which works towards interoperability and disaggregation of components in wireless networks by enabling multi-vendor deployments and software-based customization. These systems leverage a shared database that can be accessed by multiple software microservices called *xApps* that may employ ML-based architectures. There is an abundance of literature that has shown superlative results for the inclusion of ML within O-RAN with some applications being spectrum sharing [4], resource Management [5] and Cell-Free Support [6]. A recent use case of ML and O-RAN for spectrum sharing has been demonstrated in [7] where the authors developed an ML-based spectrum sensing xApp that utilizes an object detection ML model called YOLO for detecting radar signals present within the spectrograms in uplink LTE/5G communications in the CBRS band. Their purpose was to enhance the coexistence between radar systems and cellular communication by re-using existing cellular infrastructure for sensing and communication.

These ML algorithms can be developed to work on one-dimensional data like key performance metrics (KPMs) or I/Q samples by using Dense Layers [8] [9]. However, due to the nature of the inherent information, a spatio-temporal rendition like spectrograms is a superior way to use this data [10] [11], and ML models are usually developed for exploiting such a data type. Here, the data is represented as images and the obtained structural information in these spectrograms is used to develop better predictive algorithms as the resultant ML models have access to additional highly related features.

Regardless of the array of benefits these predictive pipelines show, there is a growing concern regarding the security and privacy of such data in a shared environment. *Specifically, in O-RAN architecture, there is a growing concern regarding the vulnerability of data stored in the RIC (RAN Intelligent Controller) database of the near-RT RIC.* O-RAN Alliance Working Group 11 (Security Working Group) in [12] have done a comprehensive security analysis and identified various threat models that exist in O-RAN including threat agents, threat surfaces, and threats for each O-RAN component and open interfaces.

Moreover, O-RAN Alliance Working Group 11 in [13] has identified various attack vectors and threat models that could affect ML solutions hosted as xApps in the Near-RT RIC. These threats could range from poisoning test data used by ML models to altering an ML model and breaching the confidentiality and privacy of user data.

Due to the open, shared, multi-vendor O-RAN architecture, it is paramount that these RAN data be it in the form of I/Q samples, spectrograms, or KPMs are secured against malicious adversaries that may try to exploit the data to get sensitive information on UE identity or location, traffic patterns and so on. These data can be used to identify different traffic scenarios and can be exploited by an adversary to track several UE traffic patterns which can further be used to perform things like intelligent jamming to disrupt the use of the spectrum by legitimate users.

In existing literature, most existing works on ML data privacy focus on federated learning [14]. This would require training models in a distributed manner to protect sensitive data and it primarily addresses the training information and not privacy during real-time inference. Some other works on privacy involve the use of SGX (Software Guard Extensions) for cloud [15] and secure multi-party computation [16]. These works do not address the aforementioned vulnerabilities that are prevalent in ML use cases. To the best of our knowledge, one paper that demonstrates privacy preservation is [17]. This paper leveraged MNIST and ImageNet datasets and then presented a solution using AES encryption and random permutation. The authors then obtained a final performance in the range of $10-20\%$ for fully encrypted images [17].

**Contributions.** Based on these identified vulnerabilities, our focus in this paper is securing the data obtained from the network. Specifically, our focus in this paper is on the spectrogram data type stored in the RIC database of the near-RT RIC. Moreover, we focus on securing the data that are used by the ML-driven microservices called xApps. These xApps which are located in the near-RT RIC are used for various RAN control applications, such as spectrum sharing. We propose the use of Vision Transformers due to their intrinsic ability to perceive the proposed data manipulations [18] and their robustness to the aforementioned CNN drawbacks [19].

As all the inferences are conducted on encrypted data, and no decryption key is shared to obtain the ML model's predictions, this is analogous to the current work on fully homomorphic encryption (FHE) [20]. While in FHE-driven ML the privacy is preserved but it is highly inefficient and leads to substantially larger computational times [21], our proposed system does not involve an FHE-based inference and can be implemented through the available prominent deep learning packages and is compliant with the strict latency requirements of O-RAN systems. The prominent contributions of this paper can be summarized as follows:

- For the first time, we investigate data privacy concerns within O-RAN networks. O-RAN is an open, shared, multistakeholder architecture where sensitive RAN data are stored in a shared database within the near-RT RIC which is accessible by various third-party ML-based microservices called xApps. These xApps use these data for diverse RAN control applications, such as, spectrum sharing and interference mitigation.
- Our proposed privacy-preserving solution adopts a two-step approach. First, it employs a novel shuffling-based learnable encryption to encrypt the spectrogram before storing it in the RAN database within the near-RT RIC. Subsequently, a customized vision transformer (ViT) architecture is utilized to derive predictions/inferences from the *encrypted* data, all while maintaining a reduced model size for faster prediction times. This design aligns with stringent latency requirements of near-RT RIC ranging from 10ms - 1s. This innovative approach not only safeguards data privacy but also fortifies ML-based xApps against potential threats such as model inversion and data extraction attacks.
- Leveraging an over-the-air (OTA) O-RAN testbed, we conduct an extensive analysis and comparisons of our proposed approach against state-of-the-art baselines, including CNN, ResNet, and DenseNet. Our experiments unequivocally showcase that our approach strikes a remarkable equilibrium between model accuracy and prediction times, when applied to encrypted data, thereby ensuring robust data privacy.

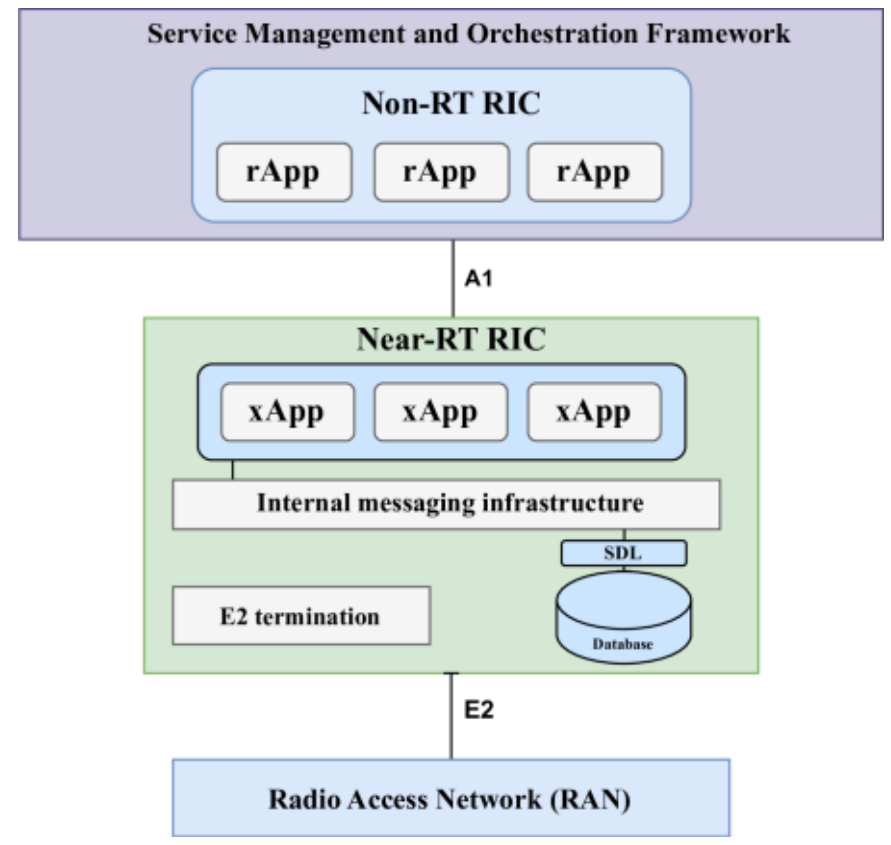

Fig. 1: Simplified O-RAN Architecture

These findings collectively highlight the promising potential of our proposed approach as a robust solution for ensuring data privacy within O-RAN systems without compromising on the model performance and stringent real-time latency requirements of O-RAN systems.

## II. O-RAN BACKGROUND

Figure 1 shows key components of an O-RAN system. An extensive explanation of the architecture and functions has been given in [22]. For our study on data privacy, we focus majorly on the near-RT RIC component which hosts other integral components we discuss briefly below.

The near-RT RIC hosts third-party vendor applications called **xApps**. These xApps act as *intelligent components* and run ML algorithms that are used to determine control policies for optimizing the RAN through the E2 interface. Other major components of the near-RT RIC include the RIC database that represents the shared data storage of the O-RAN system and internal messaging infrastructure which helps to connect multiple xApps and also ensures message routing.

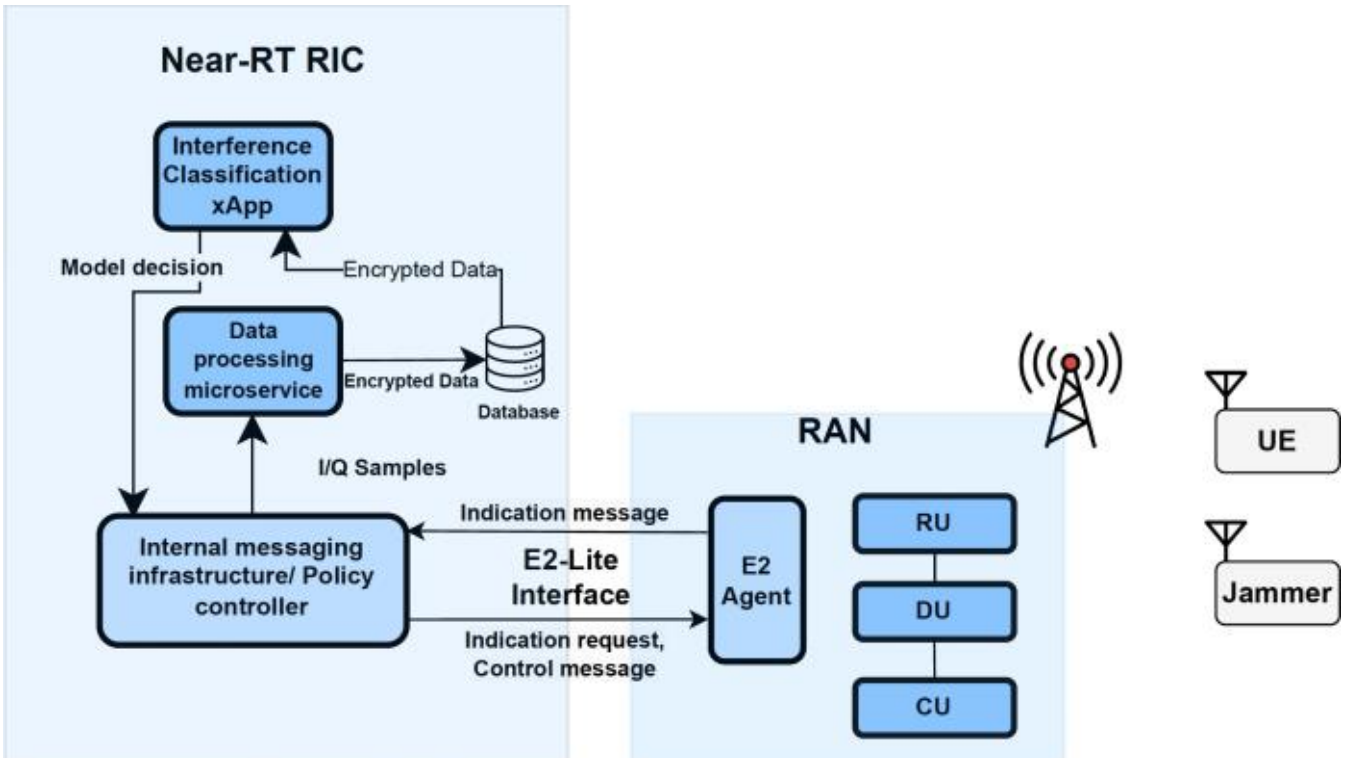

Fig. 2: Overview of O-RAN system showcasing the working of exemplary ML-based Interference Classification xApp.

### A. RIC Database

The RIC database serves as a repository for various data, including information regarding the UEs such as their location. It also contains data related to the RAN, offering insights into access network-related information influencing overall network performance. The data stored in the RIC database may encompass key performance metrics (KPMs) such as throughput and signal-to-interference-plus-noise ratio (SINR) characterizing the quality of communication between UEs and RAN. It can also include I/Q samples or spectrograms that can be used to draw several insights of the network.

The RIC database's role in data sharing fosters a collaborative ecosystem where diverse xApps can utilize these data to make informed decisions and collectively optimize network performance. This highlights its pivotal role in supporting a dynamic and agile open RAN system. Thus, it is pivotal that the data stored in this RIC database are secure and not easy to manipulate by a malicious entity.

### B. Exemplary ML-based Interference Classification xApp

For our study, we design an ML-based interference classification xApp that is used to detect the presence of jammers transmitting different kinds of interference in a network. As shown in Figure 2, it operates by taking spectrograms as input, which are stored in the RIC database within O-RAN, and makes real-time decisions about the presence of interference within the network. Upon identifying the kind of interference, the xApp sends a control message through the internal messaging infrastructure to the RAN using the E2-Lite[1] interface thus prompting it to make certain controls to optimize network performance. *The idea behind our study is to enable this ML-based xApp to be able to use the encrypted data stored in the RIC database for inference purposes.* The data processing microservice performs the conversion from I/Q samples into spectrograms and also the encryption. All the machine learning models that are utilized in this paper are deployed in this xApp and a more accurate model would result in better decisions for RAN control and a better network performance.

[1]E2-lite is a lightweight implementation of O-RAN E2 interface that enables communication between E2 nodes (i.e., RAN) and Near-RT RIC.

## III. Proposed Approach

We propose a privacy-preserving (data pipeline) methodology for a shared, multi-stakeholder O-RAN environment while emphasizing a trade-off for predictive accuracy and computational efficiency. O-RAN mandates a 10ms - 1s latency requirement for the closed-loop communication between the near-RT RIC and the RAN. The proposed approach includes two subsequent steps – (1) *Shuffling-based Learnable Encryption* that encrypts the RAN data (in the form of spectrograms), followed by (2) *Vision Transformer (ViT) Model* - that can perform inferences on encrypted data, and thus, ensuring privacy preservation of RAN data in the shared RIC database in an open, multistakeholder O-RAN environment.

### A. Shuffling-based Learnable Encryption

The proposed Shuffling-based Learnable Encryption method involves the conversion of the spectrogram image to a set of $N \times N$ grids (where $N$ is a hyperparameter dependent on the predefined patch size). Specifically, the proposed technique conducts two subsequent shuffling-based operations to create an encrypted spectrogram –

1) *Grid-based Shuffling:* In this operation, $N \times N$ grids are randomly shuffled. Figure 3 (A) showcases the encrypted spectrogram after this operation, and
2) *Pixel-based Shuffling:* For each resultant grid, the pixels are shuffled and the outcome can be seen in Figure 3 (B).

As the encryption still results in a spectrogram where essentially every pixel has been shuffled through a different random seed, it is extremely difficult to retrieve the original spectrogram, hence enhancing effective privacy. If there is a malicious xApp, the malicious xApp cannot understand the shuffled or encrypted images, and it would be difficult to reverse the said process as seen in the paper [18]. Therefore, we can confidently say that privacy is preserved. The paper [18] also validated the potency of this method by subjecting this and other alternatives to a multitude of reconstruction attacks.

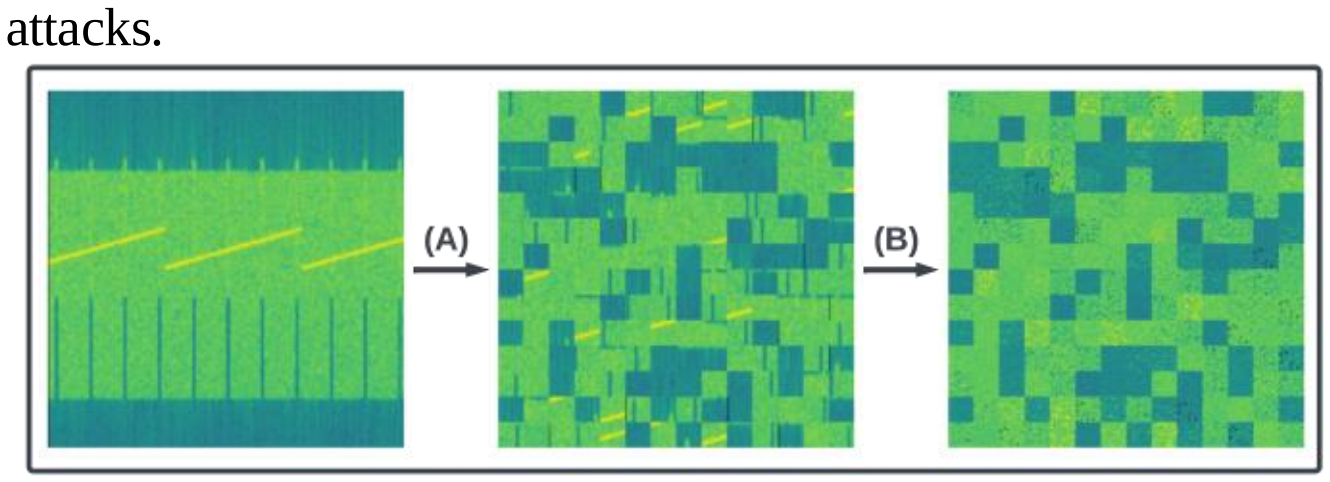

Fig. 3: Overview of Shuffling-based Learnable Encryption: (A) grid-based shuffling, and (B) pixel-based shuffling.

### B. Vision Transformer (ViT) Model

We first elaborate on the motivation behind using the ViT model as a subsequent key component of the proposed privacy-preserving computation pipeline. Following that, we give a detailed explanation of the ViT architecture.

*1) Motivation behind using ViT Model:* There are two properties of the Vision Transformer architecture that depict a superior pair-up with the encryption pipeline. As the ViT showcases a superlative invariance and robustness to patch order (the spatial location of a particular grid) patch order and its intrinsic nature of working with image patches instead of an entire image it is highly relevant for our study [18].

Patch-order invariance is the property where the output of the transformer encoder remains consistent and unaffected, regardless of the sequence or arrangement of input patches. We can also say that the pixel shuffling aspect of the encryption pipeline can be perceived by the Linear Embedding Layer due to the encryption being a learnable linear transformation the pixel shuffling aspect of the encryption pipeline should also be addressed accurately [18]. We believe that the natural affinity of the ViT model to perceive such encrypted images is an important motivation, and it should enable us to create a viable ML predictor with fewer parameters which are extremely important for a real-time network use case.

*2) ViT Architecture:* Having a shallower model is necessary for achieving the strict latency constraints and at the same time, having an acceptable level of accuracy is also important. If we look at contemporary research in ML the accurate models do showcase a larger model size, hence having an architecture that naturally supports the encryption technique does showcase substantial utility and the possibility of having a sufficiently accurate model with considerably lesser parameters. Keeping these in mind, we design a shallower ViT model, where the architecture is divided into three primary parts, the linear Embedding layer, and the transformer encoder, which is followed by the softmax-activated classifier head. The internal functioning is also depicted in Figure 4.

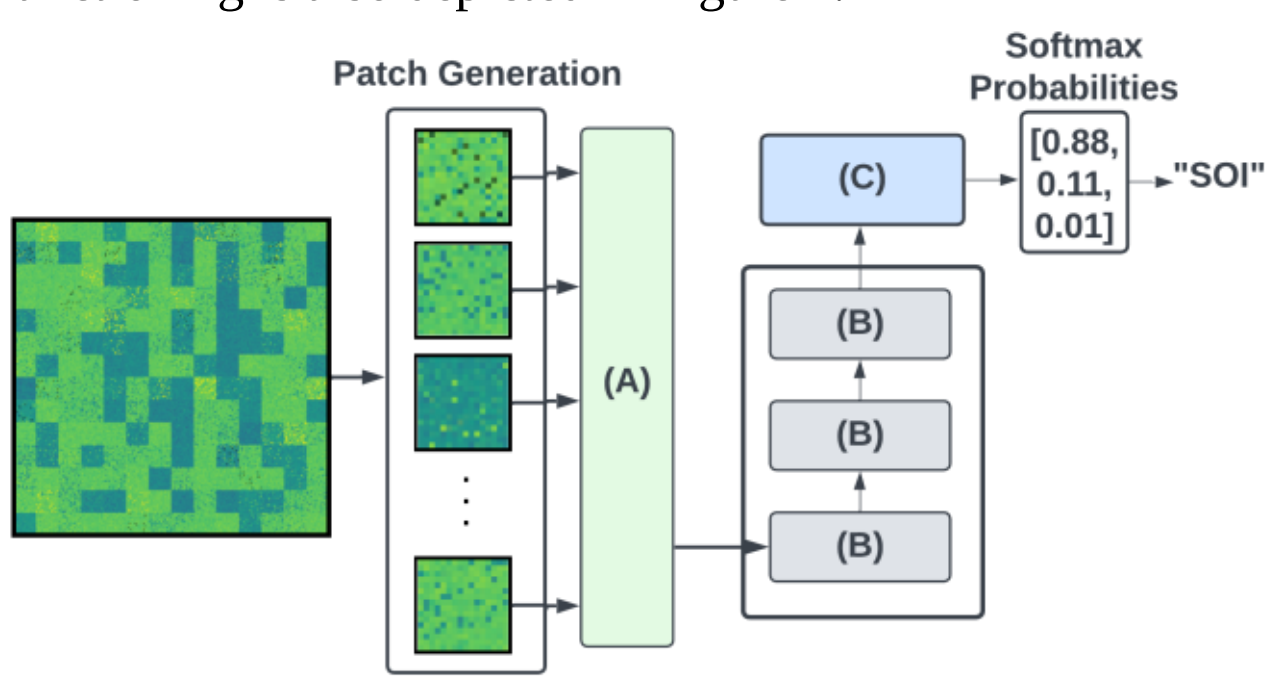

Fig. 4: Overview of ViT architecture: (A) showcases the Linear Embedding layer, (B) showcases the Transformer layers, and (C) is the classifier head.

*a) Linear Embedding Layer:* The linear embedding layer is responsible for converting the input sequence of image grids to a continuous vector representation. It leverages a learnable Embedding Matrix $E$ of dimensionality $d$ which maps each patch to a lower-dimensional vector space and a Classification Token ($v_c$) which is concatenated into the embedded image patches to mimic the original transformer architecture [23]. The classification token serves as a global representation of the entire image, allowing the model to make predictions by considering both local patch embeddings and the essential global context provided by this token. The governing equation for the linear transformation layer can be perceived as:

$$k_0 = [v_c; a_1E; a_2E; \ldots ; a_nE] + E_p \tag{1}$$

Here, $E_p \in \mathrm{R}(n + 1) \times d$ and the variable $k_0$ represents the resultant embedding sequence, $E_p$ pertains to the positional information and instills spatial information, and $a_i$ represents the image patches where the maximum value of $i$ is $n$ [23].

*b) Transformer Encoder:* The transformer encoder primarily consists of the Multi-head Self-Attention block which is abbreviated as $MSA$ and is responsible for gauging the relative importance of each patch embedding in comparison with the other embeddings, and the fully connected feed-forward dense block which can be perceived as a Multi-Layered Perceptron or $MLP$ [23] [24]. The governing equations for both use the paradigm of Layer normalization $LN$ and can be explained as:

$$k'_\ell = MSA(LN(k_{\ell-1})) + k_{\ell-1}, \quad \ell = 1 \ldots J \tag{2}$$

$$k_\ell = MLP(LN(k'_\ell)) + k'_\ell, \quad \ell = 1 \ldots J \tag{3}$$

For obtaining the softmax predictions the first element of the sequence $K_{o\ell}$ is used as an input for a final dense layer. The $MSA$ further consists of the self-attention layer, and the concatenation layer, which is responsible for combining the outputs of the multiple attention heads [23] [24]. Here, the number of dimensions for the $MLP$ output is kept at 128, the hidden size for the embedding layer is fixed at 64, and each $MSA$ has 4 attention heads. Here, the encoder would include $J$ identical layers and to obtain a lower inference time, we only use 3 transformer layers.

## IV. Results and Discussion

This section conducts a thorough analysis of the proposed approach against several baselines (see subsection IV-A) using an OTA O-RAN testbed (see subsection IV-B) with ML-based Interference classification xApp (Section III-B) as the exemplary case study for ML data pipeline within an open, shared, multistakeholder O-RAN network.

### A. Comparison Baselines

*1) Convolutional Neural Network (CNN) Architecture:* The proposed ViT model has three transformer layers which are followed by the classifier head. To have a similar parameterized model for the baseline CNN, we use three convolutional layers with 64 filters of size (3,3), where each layer is followed by the aforementioned MaxPooling layer that is finally succeeded by a flattening layer and a Softmax activated classifier for obtaining the probabilistic distribution.

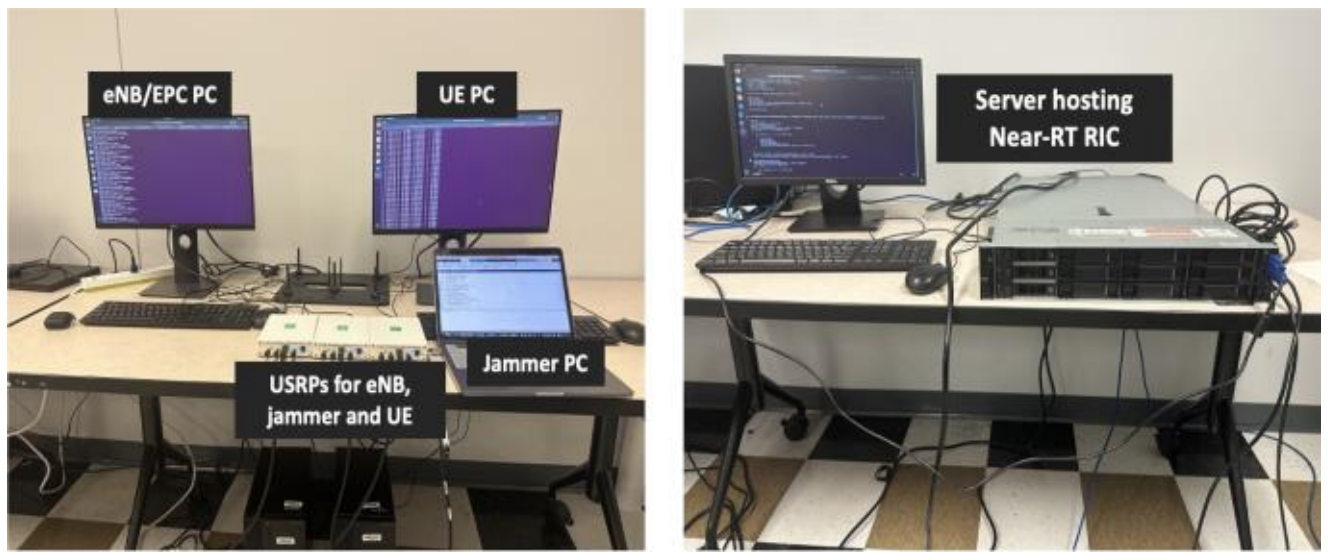

Fig. 5: O-RAN testbed. The left image shows our base station, user equipment, and the jammer USRPs. The right image shows the server hosting the near-RT RIC

*2) Other Baselines (ResNet-50, DenseNet-121, MobileNet-v2, and VGG):* We also compare the proposed approach with other prominent CNN-based architectures, mainly ResNet-50 [25], DenseNet-121 [26], MobileNet-v2 [27], and VGG [28]. The MobileNet was developed to implement CNN in resource-constrained devices with a lower inference time [27]. The ResNet is a very deep architecture that leverages skip-connections, it has shown superlative results for multiple utilities but contains significantly higher parameters. Similar to the ResNet, the DenseNet and VGG also implicate a strong baseline, however, they are very deep models with substantially higher model sizes. To implement such existing architectures, we import the models without any pre-trained weights without any classifier head and append the model with a Global Average Polling Layer [29], a hidden layer with 128 ReLU activated neurons, and a SoftMax layer for the classification probabilities. The Global Average Polling Layer is used to reduce the spatial dimensions of feature maps and encourage the model to focus on more meaningful and compact representations

### *B. Experimental Setup*

As illustrated in Figure 5, our O-RAN testbed comprises a RAN/core network which is co-located on the same computing system, a UE, and then, a jammer responsible for generating jamming/interference signals on the uplink signal of the UE. The near-RT RIC is hosted on a rack server and can serve multiple base stations. The RAN and UE are implemented using the open-source srsRAN cellular software stack (version 21.10), which is designed for building LTE/5G cellular networks [30]. We made modifications to the srsRAN codebase to tailor it for our testbed's requirements, adding functionalities like creating a buffer for storing collected I/Q samples and certain RAN control capabilities such as switching between adaptive or fixed MCS based on the decision from the interference classification xApp.

We employ USRP B210 SDRs as the RF devices for both the RAN and UE. For the near-RT RIC, we utilize the O-RAN software community's open-source codebase to implement it. We compile this source code on our server and establish connections to the base station via an E2-lite interface (this is a simple implementation of the E2 interface) [31]. The jamming signals are generated using MATLAB and transmitted over the air (OTA) using another USRP.

### *C. Dataset Generation*

The data used for our study was collected using the experimental setup described previously. Our dataset comprises 2100 spectrograms, which have been divided into three distinct classes for training purposes with 700 samples each. We also normalize all the images between the range of [0, 1] by dividing each pixel by 255. As we perform experiments on three different patch sizes, the same images are leveraged for the model training and inferences and each encrypted image is generated with a different random key, leading to a more difficult performance benchmark.

The first class represents the uplink UE signal with no interference which we call signal of interest (SOI). These SOI are transmitted at an uplink carrier frequency of 2.56 GHz. For network configurations, we leveraged 25 physical resource blocks (PRBs), which correspond to approximately 5 MHz of bandwidth, necessitating a sampling rate of 7.68 Mega samples per second. We also generated uplink TCP traffic at a rate of 5MHz between the UE and the base station using iperf3. We set up an iperf3 server at the RAN end, with the iperf3 client running on the UE side.

The second class and third class used for training data represent scenarios with interference, specifically continuous wave interference (CWI) and chirped interference (CI). These interference signals were generated at various gain values ranging from 30 dB to 40 dB. There are 700 CWI and 700 CI spectrograms respectively.

The SOI signals were transmitted leveraging the open-source srsRAN stack, while the jamming signals CWI and CI were transmitted OTA on the same carrier frequency as the SOI. We used a MATLAB-generated script to generate the baseband I/Q samples for the CWI and CI signals and then utilized another USRP to transmit these signals OTA.

### *D. ML model Setting*

The primary mode of comparison is performed via a stratified train-test split where 70% of the total data is reserved for training, 15% is used for validating the models and for selecting the best weights through multiple epochs, and the rest is reserved for testing. All the models are trained using early stopping and the 'ReduceLROnPlateau' functionality as callbacks. The maximum epochs are kept as 100 for the shallower models and 35 for the denser models to adhere to the computational constraints. All the architectures are trained on an identical data distribution to remove unwanted biases and maintain a fair comparison. We also present the associated temporal metrics.

### *E. Model Accuracy and Prediction Times*

The primary experiments (i.e., Model accuracy and Prediction time) are performed on the aforementioned train-test split, and the obtained results are mentioned in table I. We include classification metrics like Precision, Recall, F1-Score, Accuracy, as showcased in the paper [32], and inference time.

TABLE I: Results for the main experiments, here the prediction times are calculated for obtaining the predictions for a single image and are mentioned in seconds. The parameters are mentioned in the orders of Millions.

| Models | F1 Score | Accuracy | Precision | Recall | Prediction Time (s) | Parameters (M) |
|---|---|---|---|---|---|---|
| CNN (Baseline) | 0.567 | 0.565 | 0.570 | 0.570 | 0.097 | 0.205 |
| ViT (Proposed Work) | 0.806 | 0.810 | 0.807 | 0.810 | **0.094** | **0.162** |
| ResNet50 | 0.855 | **0.854** | 0.857 | 0.854 | 0.237 | 23.850 |
| DenseNet121 | 0.767 | 0.762 | 0.828 | 0.762 | 0.266 | 7.169 |
| MobileNetV2 | 0.167 | 0.333 | 0.111 | 0.333 | 0.131 | 2.422 |
| VGG16 | 0.167 | 0.333 | 0.111 | 0.333 | 0.295 | 14.780 |

From the table I, it can be inferred that the proposed approach (utilizing the ViT model) demonstrates the most favorable balance between classification metrics and ML inference time when we apply the shuffling-based encryption on our dataset. All models, excluding MobileNet and VGG-16, show convergence and valid performance scores compared to the baseline CNN. Specifically compared to baseline CNN (preferably used as a status-quo ML technique in wireless networks), our proposed ViT approach significantly outperforms the baseline CNN, achieving a remarkable absolute increase of 24.5% and 23.9% for accuracy and F1-Score, respectively. The only model that was able to outperform the ViT was the ResNet-50 architecture, with an increase of 4.4% and 4.9% for percent accuracy and F1-Score. However, *it requires a substantially large number of parameters and thus, suffers from huge inference time overheads*, which may become a bottleneck in utilizing such deep models in latency-critical O-RAN systems. The ViT is capable of performing similarly to a ResNet-50 while showcasing a 99.32% decrease in the total parameters and a 60.23% decrease in the model prediction times for one fully encrypted image. The ViT was able to showcase a better performance than the DenseNet which is also a prominent CNN-based architecture for classification tasks, with an increase of 4.8% and 3.9% for percent accuracy and F1-Score respectively.

1) *Confidence Scores:* We evaluate models at various confidence thresholds to grasp their performance across different inputs and error rates. Typically, models offer probability distributions for each class, and we commonly select the highest probability as the predicted class. To enhance interpretability, we treat each class probability as a confidence score. By incrementally adjusting the threshold, we wait for confidence scores to surpass it before making final predictions, thereby improving explainability. The pictorial representation of this experiment is mentioned below in figure 6.

As we are only registering the predictions where the maximum softmax probability exceeds a particular value, it is intuitive to believe that the number of predictions would decrease. The obtained graph further reinforces our results as the ViT significantly performs the baseline CNN both for the prediction accuracies and the data prediction rates. It was also interesting to see that the ViT has a very similar performance to the ResNet and DenseNet at substantially lesser parameters, and it was able to outperform the said approaches at stricter confidence thresholds. We have not tested the MobileNet and VGG for this due to the poor performance of the original test set and the lack of convergence.

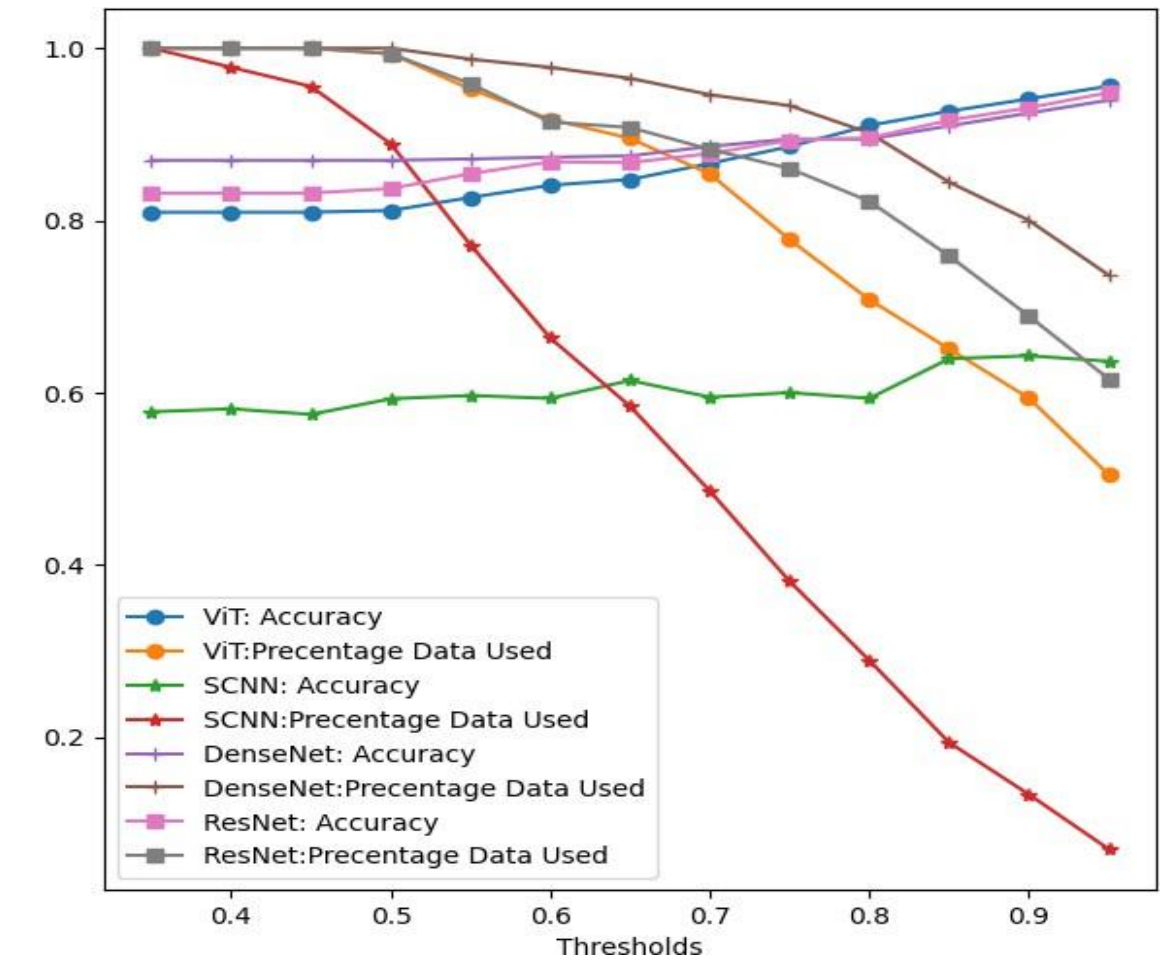

Fig. 6: Accuracy and Data usage trends for different confidence thresholds.

TABLE II: Experiments pertaining to different patch sizes

| Model | Patch Size: 8 | | Patch Size: 32 | |
|---|---|---|---|---|
| | F1 Score | Accuracy | F1 Score | Accuracy |
| CNN (Baseline) | 0.674 | 0.679 | 0.167 | 0.333 |
| ViT (Proposed Work) | 0.754 | **0.762** | 0.728 | 0.740 |
| ResNet50 | 0.735 | 0.746 | 0.856 | 0.857 |
| DenseNet121 | 0.433 | 0.543 | 0.898 | **0.898** |
| MobileNetV2 | 0.167 | 0.333 | 0.167 | 0.333 |
| VGG16 | 0.167 | 0.333 | 0.167 | 0.333 |

### *F. Hyperparameter Analysis*

This section indicates the auxiliary experiments which primarily include real-time characteristics, variable patch sizes, token sparsification, and tests for shuffling-based invariance.

1) *Variable Patch Sizes:* As the encryption techniques rely on a patch size parameter, we can alter the variable to increase the effective randomness, hence stronger encryption.

From the table II, we can further solidify the inference that ViT offers a superlative trade-off between accuracy and temporal trade-offs. It was also interesting to note that at larger patch sizes when each patch has access to more information the deeper models like ResNet and DenseNet were able to showcase a better performance. For a smaller patch size, the ViT outperformed all the other models, while maintaining a performance boost relative to the baseline at all times. The two architectures, MobileNet and VGG also presented an inferior performance for this experiment.

2) *Tests for Shuffle Invariance:* As it is possible to create multiple encrypted versions of the same spectrogram by the leveraged encryption mechanism, it is also necessary to understand how these different models perform when provided the same information with a different encrypted form. If a model can provide us the same prediction on different encrypted versions of the same image we can say that the model is invariant to the grid shuffling and it has learned specific features that are invariant to the random seed. It can also implicate that the model has learned the required features despite the structural information that happens when we shuffle the grids on a spatial level. This experiment is conducted by taking 30 new samples for each class and by generating 15 encrypted versions for each image. We also check how the model functionalities change for each class and the overall accuracy associated with the newer combined distribution. The obtained results are mentioned in figure 8.

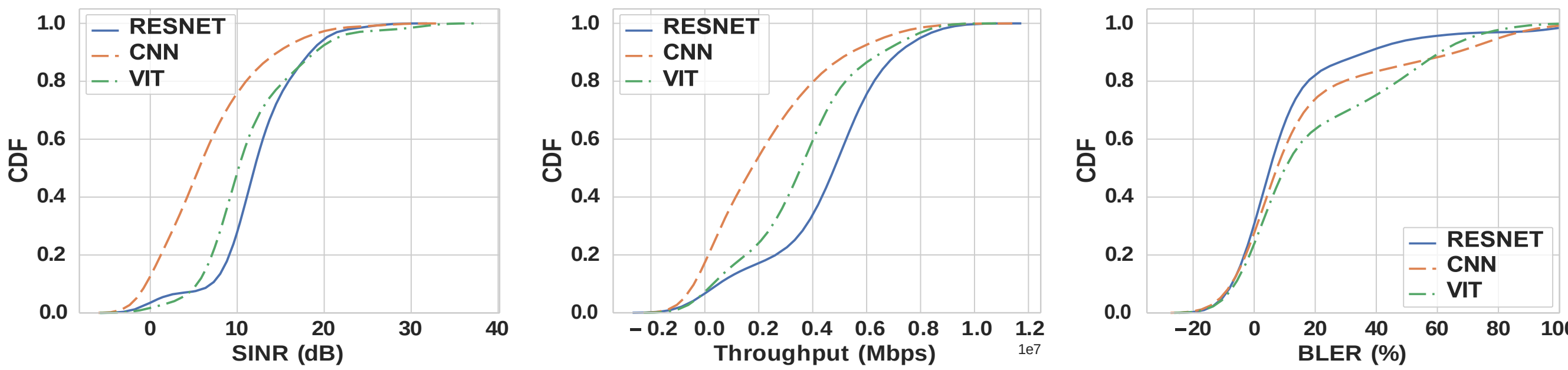

(a) (b) (c)

Fig. 7: CDF plots for (a) SINR, (b) Throughput, and (c) BLER for each considered ML model.

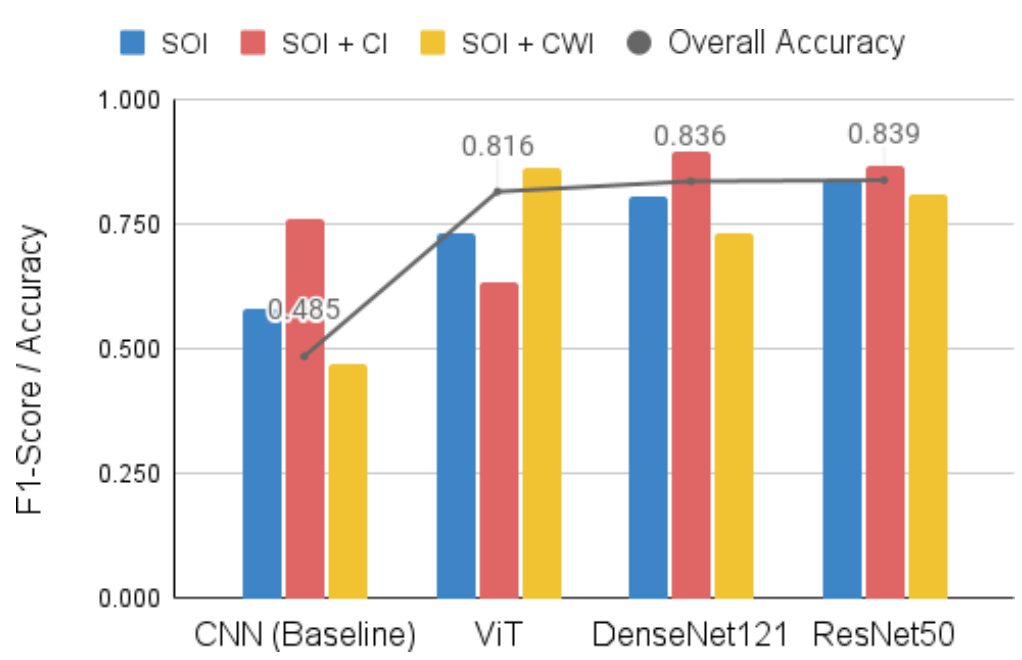

Fig. 8: Results pertaining to shuffling-based invariance.

The results reinforce the previous inferences concerning model performances. We can also see that the SOI+CI was the easier class to predict for the CNN-based models, and the ViT was able to outperform the other models for predicting the SOI+CWI. Since all models can showcase an acceptable level of robustness in relation to the standard result from table I which shows the performance metrics, we can say that the models could learn some features from the shuffled data and that this method of encryption is learnable by such deep-learning architectures.

### *G. Network Performance Evaluation*

Here, we compare the performance of our proposed ViT model against two other architectures, the baseline CNN and the ResNet-50 which were deployed as ML models in our xApp located in the near-RT RIC using the O-RAN testbed. The CNN was our primary baseline and was developed to showcase an analogous architecture to the proposed ViT. The ResNet-50 was the only convolutional-based network that gave a better performance than our ViT but required substantially more parameters.

To evaluate network performance on each model deployed, we initiate uplink traffic from the UE to the RAN, spanning a duration of 180 seconds. In the initial 90 seconds, the UE transmits uplink traffic without interference from the jammer. Subsequently, for the next 90 seconds, we deliberately introduce OTA interference from the jammer, with a relative gain of 40 dB. The plots in Figure 7 showcase the Cumulative Distribution Function (CDF) of SINR (dB) (Figure 7(a)), uplink throughput (Figure 7(b)) and Block Error Rate (BLER) (Figure 7(c)). Notably, from observation, we can see that the ResNet-50 model outperforms the ViT and CNN models in terms of network performance. This can be attributed to the fact that RESNET achieves the highest accuracy compared to the other two models (See Table 1). Nevertheless, the ViT model also demonstrates a network performance comparable to ResNet-50 with considerably fewer parameters, while CNN, on the other hand, exhibits the worst network performance, primarily attributed to its lower accuracy.

### *H. O-RAN Timing Evaluation*

To evaluate the performance of our O-RAN system in terms of latency, we have analyzed the round trip timing (RTT) it takes when each model is deployed on our testbed. This RTT encapsulates the time it takes the data to be sent from the RAN to the near-RT RIC, the time it takes to process the data and store it in the database, the time for model inference, and then the time to send control decision back to the RAN via the E2-Lite interface. All these processes can be visualized in Figure 2. For the ViT, CNN, and ResNet-50 models, the total RTT were 611.05ms, 584.4ms, and 713.13ms respectively. The ResNet-50 model has the highest RTT time which can be attributed to the fact that it has the highest number of parameters. Therefore, we can see that we are still under the latency requirement of 1s mandated by O-RAN for closed-loop control and communication between the near-RT RIC and RAN.

## V. Conclusion and Future Work

The paper proposed a privacy-preserving solution for securing RAN spectrogram data against various privacy attacks/leaks in an open, multistakeholder wireless environment with a case study of O-RAN networks. The proposed approach comprised of two subsequent steps – first it encrypts the RAN spectrogram data using a shuffling-based learnable encryption model, that can be stored in the shared environment. Following this, it utilizes a ViT model that can perform accurate inferences on encrypted data within the required latency constraints. Our extensive analysis using an OTA O-RAN testbed against various baselines demonstrated the superiority of the proposed approach in achieving better inference accuracy as well as lower prediction times while ensuring data privacy, which together make it suitable for deployments for latency-critical ML-driven applications in wireless networks. Looking ahead, our future research aims to explore deeper transformer architectures and other speed-up techniques that can be leveraged to incorporate more intricate and parameterized architectures while accounting for compliance with the established wireless network standards. As we were able to see promising results from the ViT model by only using 3 transformer layers, we also wish to work towards deeper architectures as it can be assumed that an equally deep and parametrized model can further enhance the performance significantly.